\documentclass[useAMS,usenatbib]{mn2e}

\usepackage{graphicx}
\bibpunct{(}{)}{;}{a}{}{,}
\usepackage{txfonts}

\newcommand{\ion}[2]{#1\,{\sc #2}}

\title[Magnetic field of HD\,345439]{
Characterizing the magnetic field and spectral variability of the rigidly rotating magnetosphere star
HD\,345439
}

\author[Hubrig et al.\ 2015]{
S.~Hubrig$^1$\thanks{E-mail: shubrig@aip.de},
A.~F.~Kholtygin$^2$, M.~Sch\"oller$^3$, I.~Ilyin$^1$\\
$^1$ Leibniz-Institut f\"ur Astrophysik Potsdam (AIP), An der Sternwarte~16, 14482~Potsdam, Germany\\
$^2$ Saint-Petersburg State University, Universitetskij pr.~28, Saint-Petersburg 198504, Russia\\
$^3$ European Southern Observatory, Karl-Schwarzschild-Str.~2, 85748~Garching, Germany
}

\begin{document}

\date{Accepted Received; in original form}

\pagerange{\pageref{firstpage}--\pageref{lastpage}} \pubyear{2015}

\maketitle

\label{firstpage}

\begin{abstract}
A team involved in APOGEE, one of the Sloan Digital Sky Survey~III programs, recently announced the 
discovery of two rare rigidly rotating magnetosphere stars, HD\,345439 and HD\,23478.
Near-infrared spectra of these
objects revealed emission-line behavior identical to that previously discovered in the
helium-strong star $\sigma$\,Ori\,E, which has a strong magnetic field and rotates fast.
A single spectropolarimetric observation of HD\,345439 with FORS\,2 at the VLT in 2014 over 88\,min indicated that 
HD\,345439 may host a strong, rapidly varying magnetic field. 
In this work, we present the results of our spectropolarimetric monitoring of this star with FORS\,2,
which revealed the presence of a strong longitudinal magnetic field
dominated by a dipolar component. The analysis of spectral variability indicates an opposite behaviour
of He and Si lines, which is usually attributed to differences in the distribution of surface He and Si  
abundance spots. 

\end{abstract}

\begin{keywords}
stars: early type ---
stars: individual: HD\,345439 ---
stars: magnetic field ---
stars: chemically peculiar ---
stars: abundances ---
techniques: polarimetric
\end{keywords}

\section{Introduction}
The presence of a rigidly rotating magnetosphere in the early B-type stars HD\,23478 and HD\,345439 
was recently discovered 
in the Apache Point Observatory Galactic Evolution Experiment (APOGEE; \citealt{Eikenberry2014})
using high-resolution ($R\sim22\,500$) near-infrared H-band spectra. The authors detected in the
APOGEE bandpass prominent Brackett series emission lines with a characteristic double-horned profile. The type
of profile and peak separation is typical for the rigidly rotating magnetosphere (RRM, \citealt{Townsend2005}) 
feature previously 
discovered in the fast rotating helium-strong star $\sigma$\,Ori\,E,
which possesses an extremely large magnetic field (e.g., \citealt{Landstreet1978,Oksala2015}). 
Such stars are extremely rare: the discovery of HD\,23478 and HD\,345439 has enhanced the number 
of ``extreme'' rotators by 50\% \citep{Eikenberry2014}.
The authors reported that the optical spectra of HD\,345439 reveal strong \ion{He}{i} lines and very fast 
rotation of $\sim270\pm20$\,km\,s$^{-1}$
Subsequently, \citet{Wisn2015} analysed multi-epoch photometric observations 
of this star from the Kilodegree Extremely Little Telescope, Wide Angle Search for Planets, and ASAS 
surveys revealing the presence of a $\sim$0.7701\,day rotation period in each data set. The authors suggest
that the He-strong star HD\,345439 of spectral type B2\,V
is among the faster known He-strong $\sigma$\,Ori\,E analogs, HR\,7355 ($P_{\rm rot}=0.52\,d$ -- \citealt{Rivinius2013}) and
HR\,5907 ($P_{\rm rot}=0.51\,d$ -- \citealt{Grunhut2012}). 

\citet{Hubrig2015a} carried out a spectropolarimetric follow-up of HD\,345439 on one occasion, obtaining eight subexposures
over 88\,minutes in 2014 June with the 
FOcal Reducer low dispersion 
Spectrograph (FORS\,2; \citealt{Appenzeller1998}) mounted on the 8\,m Antu telescope of the VLT.
The authors reported that the mean longitudinal magnetic field was changing from about $+$500\,G measured in the first
pair of subexposures to about $-$1200\,G measured in the last pair of subexposures.
Multi-epoch FORS\,2 spectropolarimetric observations distributed over about two months were recently 
obtained in service mode in the framework of our programme 097.D-0428.
In the following sections, we present the results of our magnetic fields measurements, the search for 
a magnetic field periodicity, and discuss the detected spectral variability with respect to the magnetic 
field geometry.

\section{Observations and magnetic field measurements}
\label{sect:obs}

\begin{table*}
\caption{
Logbook of the FORS\,2 polarimetric observations of HD\,345439, including 
the modified Julian date of mid-exposure followed by the
achieved signal-to-noise ratio in the Stokes~$I$ spectra around 5200\,\AA{},
and the measurements of the mean longitudinal magnetic field using the 
Monte Carlo bootstrapping test, for the hydrogen lines and all lines.
In the last columns, we present the results of our measurements using the null spectra for the set
of all lines and the phases calculated
relative to the zero phase corresponding to a positive field extremum at MJD56925.5425.
All quoted errors are 1$\sigma$ uncertainties. 
}
\label{tab:log_meas}
\centering
\begin{tabular}{lrr@{$\pm$}rr@{$\pm$}rr@{$\pm$}rr}
\hline
\hline
\multicolumn{1}{c}{MJD} &
\multicolumn{1}{c}{SNR$_{5200}$} &
\multicolumn{2}{c}{$\left< B_{\rm z}\right>_{\rm hydr}$} &
\multicolumn{2}{c}{$\left< B_{\rm z}\right>_{\rm all}$} &
\multicolumn{2}{c}{$\left< B_{\rm z}\right>_{\rm N}$} &
\multicolumn{1}{c}{Phase}\\
 &
 &
 \multicolumn{2}{c}{[G]} &
 \multicolumn{2}{c}{[G]}  &
 \multicolumn{2}{c}{[G]} &
 \\
\hline
     56810.2547&  413 &   414  & 282 &   436  & 212 & \multicolumn{2}{c}{} & 0.311 \\ 
     56810.2745&  455 &   789  & 246 &   565  & 188 & \multicolumn{2}{c}{} & 0.336\\  
     56810.2860&  392 & $-$303 & 282 &$-$298  & 212 & \multicolumn{2}{c}{} & 0.352  \\ 
     56810.3018&  420 & $-$840 & 262 &$-$689  & 198 & \multicolumn{2}{c}{} & 0.372 \\ 
     57525.2797&  521 &  1202  & 552 &   1310 & 374 &$-$275 & 223& 0.697\\  
     57527.2841&  729 &   287  & 302 &    416 & 206 & $-$29 &166 & 0.300\\ 
     57530.3230&  960 &  1714  & 245 &  1237  & 186 &$-$181 &122 & 0.246\\  
     57530.3753& 1086 &   514  & 185 &   518  & 141 &  $-$5 &104 & 0.314\\  
     57531.2763&  756 & $-$829 & 408 & $-$475 & 222 &$-$274 &274 & 0.483\\  
     57531.3233&  811 & $-$103 & 371 & $-$576 & 203 &    76 &205 & 0.544 \\  
     57534.3069&  786 & $-$853 & 280 & $-$926 & 181 &   123 &194 & 0.418 \\
     57560.1708&  881 &   3415 & 344 &  3044  & 235 &   113 &178 & 0.000 \\
     57590.1750& 1127 &   2546 & 184 &  2551  & 121 &   130 & 97 & 0.957 \\
     57590.2287& 1174 &   1905 & 200 &   2176 & 129 & $-$98 & 95 & 0.027 \\
     57590.2812& 1056 &   2084 & 292 &   2156 & 169 & $-$45 &113 & 0.095 \\
     57591.1437& 1053 &   1344 & 265 &   1280 & 173 &  $-$6 &145 & 0.215 \\
     57591.1997& 1178 &    826 & 199 &    583 & 137 & $-$21 & 91 & 0.288 \\
     57591.2521& 1133 & $-$372 & 229 & $-$314 & 149 & $-$51 &115 & 0.356 \\
\hline
\end{tabular}
\end{table*}

Fourteen FORS\,2 spectropolarimetric observations of HD\,345439 were obtained
from 2016 May 17 to 2016 July 22.
The FORS\,2 multi-mode instrument is equipped with polarisation analysing optics
comprising super-achromatic half-wave and quarter-wave phase retarder plates,
and a Wollaston prism with a beam divergence of 22$\arcsec$ in standard
resolution mode. 
We used the GRISM 600B and the narrowest available slit width
of 0$\farcs$4 to obtain a spectral resolving power of $R\sim2000$.
The observed spectral range from 3250 to 6215\,\AA{} includes all Balmer lines,
apart from H$\alpha$, and numerous helium lines.
For the observations, we used a non-standard readout mode with low 
gain (200kHz,1$\times$1,low), which provides a broader dynamic range, hence 
allowed us to reach a higher signal-to-noise ratio (SNR) in the individual spectra.
The exposure time for each subexposure 
accounted for 7.8\,min.

A first description of the assessment of longitudinal magnetic field
measurements using FORS\,1/2 spectropolarimetric observations was presented 
in our previous work (e.g.\ \citealt{Hubrig2004a,Hubrig2004b}, 
and references therein).
To minimize the cross-talk effect,
and to cancel errors from 
different transmission properties of the two polarised beams,
a sequence of subexposures at the retarder
position angles
$-$45$^{\circ}$$+$45$^{\circ}$,
$+$45$^{\circ}$$-$45$^{\circ}$,
$-$45$^{\circ}$$+$45$^{\circ}$,
etc.\ is usually executed during the observations. Moreover, the reversal of the quarter wave 
plate compensates for fixed errors in the relative wavelength calibrations of the two
polarised spectra.
According to the FORS User Manual, the $V/I$ spectrum is calculated using:

\begin{equation}
\frac{V}{I} = \frac{1}{2} \left\{ 
\left( \frac{f^{\rm o} - f^{\rm e}}{f^{\rm o} + f^{\rm e}} \right)_{-45^{\circ}} -
\left( \frac{f^{\rm o} - f^{\rm e}}{f^{\rm o} + f^{\rm e}} \right)_{+45^{\circ}} \right\}
\end{equation}
where $+45^{\circ}$ and $-45^{\circ}$ indicate the position angle of the
retarder waveplate and $f^{\rm o}$ and $f^{\rm e}$ are the ordinary and
extraordinary beams, respectively. 
Rectification of the $V/I$ spectra was
performed in the way described by \citet{Hubrig2014}.
Null profiles, $N$, are calculated as pairwise differences from all available 
$V$ profiles.  From these, 3$\sigma$-outliers are identified and used to clip 
the $V$ profiles.  This removes spurious signals, which mostly come from cosmic
rays, and also reduces the noise. A full description of the updated data 
reduction and analysis will be presented in a separate paper (Sch\"oller et 
al., in preparation, see also \citealt{Hubrig2014}).
The mean longitudinal magnetic field, $\left< B_{\rm z}\right>$, is 
measured on the rectified and clipped spectra based on the relation 
following the method suggested by \citet{Angel1970}
\begin{eqnarray} 
\frac{V}{I} = -\frac{g_{\rm eff}\, e \,\lambda^2}{4\pi\,m_{\rm e}\,c^2}\,
\frac{1}{I}\,\frac{{\rm d}I}{{\rm d}\lambda} \left<B_{\rm z}\right>\, ,
\label{eqn:vi}
\end{eqnarray} 

\noindent 
where $V$ is the Stokes parameter that measures the circular polarization, $I$
is the intensity in the unpolarized spectrum, $g_{\rm eff}$ is the effective
Land\'e factor, $e$ is the electron charge, $\lambda$ is the wavelength,
$m_{\rm e}$ is the electron mass, $c$ is the speed of light, 
${{\rm d}I/{\rm d}\lambda}$ is the wavelength derivative of Stokes~$I$, and 
$\left<B_{\rm z}\right>$ is the mean longitudinal (line-of-sight) magnetic field.

The longitudinal magnetic field was measured in two ways: using the entire spectrum
including all available lines, excluding lines in emission, or using exclusively hydrogen lines.
Furthermore, we have carried out Monte Carlo bootstrapping tests. 
These are most often applied with the purpose of deriving robust estimates of standard errors. 
The measurement uncertainties obtained before and after the Monte Carlo bootstrapping tests were found to be 
in close agreement, indicating the absence of reduction flaws. 

Since the presence of $\beta$~Cep-like pulsations is frequently found in early B-type stars,
we also checked  the stability 
of the spectral lines along full sequences of sub-exposures. We have compared 
the profiles of several spectral lines recorded in the parallel beam with the retarder waveplate 
at $+45^{\circ}$. The same was done for spectral lines recorded in the perpendicular beam. 
The line profiles looked identical within the noise.

The results of our magnetic field measurements, those for the entire spectrum
or only the hydrogen lines are presented in 
Table~\ref{tab:log_meas}, where we also include in the first four rows the information about the previous 
magnetic field measurements presented by \citet{Hubrig2015a}.  A non-detection was obtained by the authors, if all 
four consecutive observations recorded as pairs of position angles 
separated by 90$^{\circ}$ were combined. On the other hand, after splitting the observations into two 
data sets, i.e.\ using the first two pairs and the second two pairs consisting 
of observations at the retarder waveplate positions ($-45^{\circ}, +45^{\circ}, +45^{\circ}, -45^{\circ}$),
they obtained 3.0 to 3.8$\sigma$ detections, but with $\left< B_{\rm z} \right>$ values with opposite
sign, indicating a very fast rotation of HD\,345439. The measurements in Table~\ref{tab:log_meas} refer
to observations at just two position angles with a time lap of 22\,min. In this case,
the null profile cannot be extracted. 
The rotation phase presented in the last column of Table~\ref{tab:log_meas} was calculated assuming a period 
of 0.77018\,d, which was determined from our period search described in Sect.~\ref{sect:mag}.

\section{Period determination from the magnetic data}
\label{sect:mag}

\begin{figure}
\centering
\includegraphics[width=0.45\textwidth]{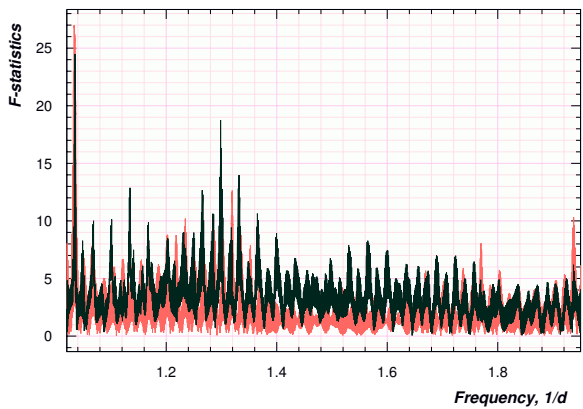}
\caption{
Frequency periodogram (in d$^{-1}$) for the longitudinal magnetic field measurements of HD\,345439 using both
the entire spectrum and only the hydrogen lines. The window function is indicated by the red color. 
}
\label{fig:period}
\end{figure}

\begin{figure}
\centering
\includegraphics[width=0.45\textwidth]{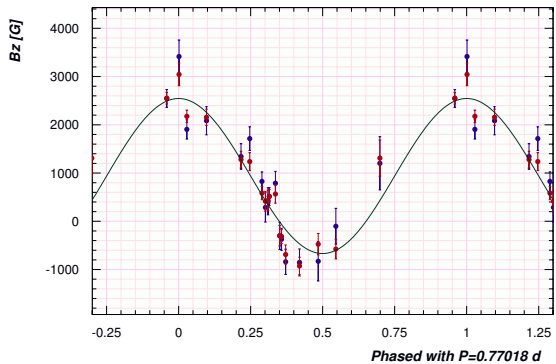}
\caption{Longitudinal magnetic field variation of HD\,345439 phased with the 0.77018\,d period.
Measurements of the magnetic field using the entire spectrum are presented by red circles,
and those using hydrogen lines by blue circles.
The solid line represents a fit to the data with a mean value for the magnetic field of
$\overline{\left<B_{\rm z}\right>} = 939\pm96$\,G and an amplitude of
$A_{\left<B_{\rm z}\right>} = 1607\pm99$\,G.
For the presented fit, we assume a zero phase corresponding to a positive field extremum at MJD56925.5425.
}
\label{fig:rot}
\end{figure}

Magnetic early B-type stars usually exhibit photometric, spectral,  and
magnetic variability with the rotation period (e.g.\ \citealt{Landstreet1978}). 
Using Kilodegree Extremely Little Telescope (KELT) 
photometry, \citet{Wisn2015} detected a rotation period of $0.7701\pm0.0003$\,d with the zero point
of the ephemeris $JD=2454252.3432$ corresponding to the first minimum in the light curve (see their
Fig.~1.). From archival photometric observations in the Wide Angle Search for Planets 
(SuperWASP) survey and the All Sky Automated Survey (ASAS),
the authors derived $P_{\rm rot}=0.7695\pm0.0078$\,d and $P_{\rm rot}=0.7702\pm0.0001$\,d,
respectively, with the phase-folded light curves exhibiting similar complex morphology. 

The result of our frequency analysis based on the longitudinal magnetic field measurements
presented in Table~\ref{tab:log_meas} and 
performed using a non-linear least squares fit to the multiple harmonics utilizing the Levenberg-Marquardt
method \citep{Press1992} with an optional possibility of pre-whitening the trial harmonics is presented in 
Fig.~\ref{fig:period}.
Since the results of the measurements using the whole spectrum or exclusively the hydrogen lines are rather similar, 
the frequency analysis 
was performed using both, the measurements on the entire spectrum and on the hydrogen lines.
To detect the most probable period, we calculated the frequency spectrum and for each trial
frequency we performed a statistical F-test of the null hypothesis for the absence of periodicity 
\citep{Seber1977}. The resulting F-statistics can be thought of as the total sum including covariances of the ratio 
of harmonic amplitudes to their standard deviations, i.e.\ a signal-to-noise ratio.
The highest peak in the periodogram not coinciding with the window function
is detected at a frequency of 1.298\,d$^{-1}$.
Using this value as an initial guess for a least-squares fit of the period,
we obtain a value of $0.77018\pm0.00002$\,d.
This  period is in good agreement with 
the results of the period search by \citet{Wisn2015} using photometric data.

In Fig.~\ref{fig:rot}, we present all measurements, those using the entire spectrum and those using only the
hydrogen lines, phased with the rotation period  and the best sinusoidal fit calculated for these 
measurements. The largest gap in the phase coverage occurs in the phase range between 0.70 and 0.95.
From the sinusoidal fit to our data, we obtain 
a mean value for the variable longitudinal magnetic field
$\overline{\left< B_{\rm z}\right>}= 939\pm96$\,G, an amplitude of the field variation 
$A_{\left< B_{\rm z}\right>}=1607\pm99$\,G, and a reduced $\chi^2$ value of 3.1. For the presented fit, we assume a zero 
phase corresponding to a positive 
field extremum at MJD56925.5425$\pm$0.0015.
The observed sinusoidal modulation indicates that the magnetic field structure
exhibits two poles and a symmetry axis, tilted with respect to the rotation axis.
The simplest model for this magnetic field geometry is based on the assumption that the studied stars 
are oblique dipole rotators,
i.e., their magnetic field can be approximated by a dipole with its magnetic axis 
inclined to the rotation axis.
In Fig.~\ref{fig:rot}, we observe around the rotational phase 0.4 noticeable 
deviations of our measurements from the simple dipole model, which may  
indicate a more complex topology of the magnetic field structure.
On the other hand, as we show later in Sect.~\ref{sect:var}, the largest dispersion in 
the hydrogen equivalent width measurements appears around the same phase range and is most likely due to an 
occultation by circumstellar gas clouds magnetically confined to the magnetic equatorial
plane (e.g.\ \citealt{Hubrig2015b}). 
Using the estimate of the stellar radius $R= 4.3\pm 0.3\,R_\odot$ for a B2\,V type star \citep{Harmanez1988},
$v \sin i = 270\pm 20$\,km\,s$^{-1}$ \citep{Wisn2015}, and
the rotation period $P_{\rm rot} = 0.77018\pm0.00002$\,d,  
we obtain $v_{\rm eq}=283\pm20$\,km\,s$^{-1}$ and an inclination angle of the stellar rotation axis to the line of
sight $i=73\pm19^{\circ}$.
From the variation of the phase curve for the
field measurements with a mean of 
$\overline{\left< B_{\rm z}\right>} = 939\pm 96$\,G and an amplitude of
$A_{\left< B_{\rm z}\right>} = 1607 \pm 99$\,G, we calculate
$\left< B_{\rm z} \right>^{\rm min}= -669\pm139$\,G and
$\left< B_{\rm z} \right>^{\rm max}=2545\pm139$\,G.
Using the definition by \citet{Preston1967}

\begin{equation}
r = \frac{\left< B_{\rm z}\right>^{\rm min}}{\left< B_{\rm z}\right>^{\rm max}}
  = \frac{\cos \beta \cos i - \sin \beta \sin i}{\cos \beta \cos i + \sin \beta
\sin i},
\end{equation}

\noindent
we find 
$r=-0.263\pm0.05$ and finally following

\begin{equation}
\beta =  \arctan \left[ \left( \frac{1-r}{1+r} \right) \cot i \right],
\label{eqn:4}
\end{equation}

\noindent
we calculate the magnetic obliquity angle $\beta=28\pm28^{\circ}$.

We can estimate the dipole strength of HD\,345439 following 
the model by \citet{Stibbs1950} as formulated by \citet{Preston1967}:

\begin{eqnarray}
B_{\rm d} & = & \left< B_{\rm z}\right>^{\rm max}  \left( \frac{15+u}{20(3-u)} (\cos \beta \cos i + \sin \beta \sin i) \right)^{-1}\\
 & \ge & \left< B_{\rm z}\right>^{\rm max}  \left( \frac{15+u}{20(3-u)}\right)^{-1}.
\end{eqnarray}

Assuming a limb-darkening coefficient of 0.3, typical for the spectral type B2V \citep{Claret2011},
we can give a lower limit for the dipole strength of $B_d \ge 8.98\pm0.49$\,kG. 

Given the high inclination angle $i=73\pm19^{\circ}$ and the low inferred obliquity angle 
$\beta=28\pm28^{\circ}$, both with rather large errors,
the estimation of the dipole strength is rather uncertain,
leading to $12.7^{+15.0}_{-3.7}$\,kG.

\section{Spectral variability}
\label{sect:var}

\citet{Wisn2015} studied the variability of several  \ion{He}{i} lines, the H$\alpha$ line and two Brackett 
hydrogen lines in the near-infrared (NIR). Although their optical and NIR spectroscopy
did not cover the full rotation cycle, the temporary changes in line profiles showed a clear correlation with
the rotational phase. 
As FORS\,2 spectra have a much lower spectral resolution, we were able to carry out a variability study
using only the strongest lines belonging to three elements: hydrogen, helium, and silicon. 

\begin{figure}
\centering
\includegraphics[width=0.47\textwidth]{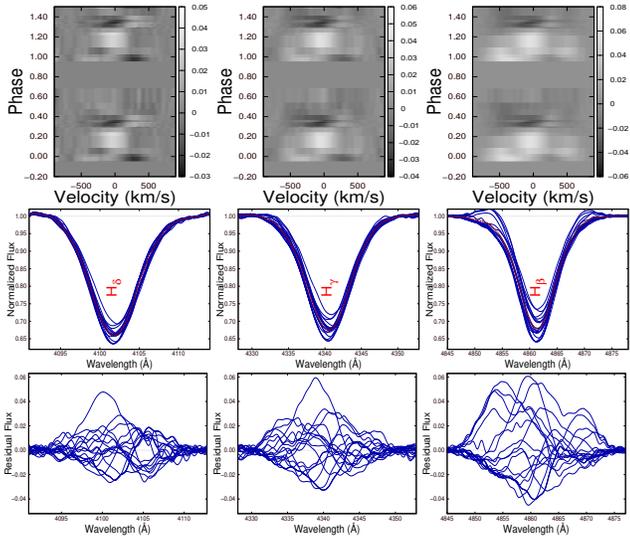}
\caption{
Variability of hydrogen lines in the FORS\,2 spectra of HD\,345439 over the rotational cycle.
The middle and lower panels show the overplotted profiles of the hydrogen lines
H$\delta$, H$\gamma$, and H$\beta$
and the differences between individual and the average line profiles. The upper panel presents the 
temporal behaviour of these differences. The average profile is shown by the red line.
}
\label{fig:hydr}
\end{figure}

\begin{figure}
\centering
\includegraphics[width=0.47\textwidth]{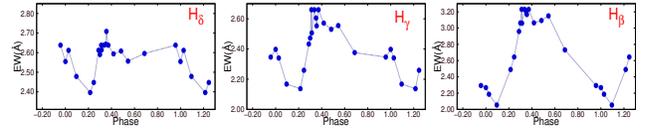}
\caption{
Variability of EWs of hydrogen lines in FORS\,2 spectra of HD\,345439 obtained at eighteen different rotational phases.
}
\label{fig:ew_hydr}
\end{figure}

In Fig.~\ref{fig:hydr}, we present the overplotted profiles of the hydrogen lines
H$\delta$, H$\gamma$, and H$\beta$,
the differences between the individual and average profiles, and 
the temporal behaviour of these differences in differential dynamic plots.
Significant emission in the wings of the hydrogen lines, best visible in the differential dynamic plot 
of H$\beta$, appears at the rotational phase around zero,
which corresponds to the maximum of the positive magnetic field strength.
Notably, we observe in the H$\beta$ emission wings a slight asymmetry, i.e.\ the 
blue-shifted emission is somewhat stronger and more extended than the redshifted emission. This behaviour 
of the H$\beta$ line differs from the behaviour of the H$\alpha$, Br$11$,  and Br$\gamma$ lines presented by 
\citet{Wisn2015} in the phase range 0.86--0.18, indicating a decrease of the blue-shifted emission with increasing
wavelength. The phase range 0.86--0.18 was calculated taking into account the difference
in the adopted zero points of ephemeris between the work of \citet{Wisn2015} and our work.
In Fig.~\ref{fig:ew_hydr}, we present the variability of the equivalent widths (EWs) of
hydrogen absorption lines showing a minimum at rotational phase 0.1-0.2, which is slightly offset from 
the positive magnetic pole, and a secondary less pronounced minimum close to the negative magnetic pole. 
The presence of intensity minima at these phases is likely related to the stronger hydrogen line profile fill-in 
by the  emission presumably originating in
the corrotating magnetospheric clouds (e.g.\ \citealt{Sundqvist2012,Hubrig2015b}).
As already mentioned in Sect.~\ref{sect:mag}, a large dispersion of EW measurements is detected around the phase 0.4.

\begin{figure}
\centering
\includegraphics[width=0.47\textwidth]{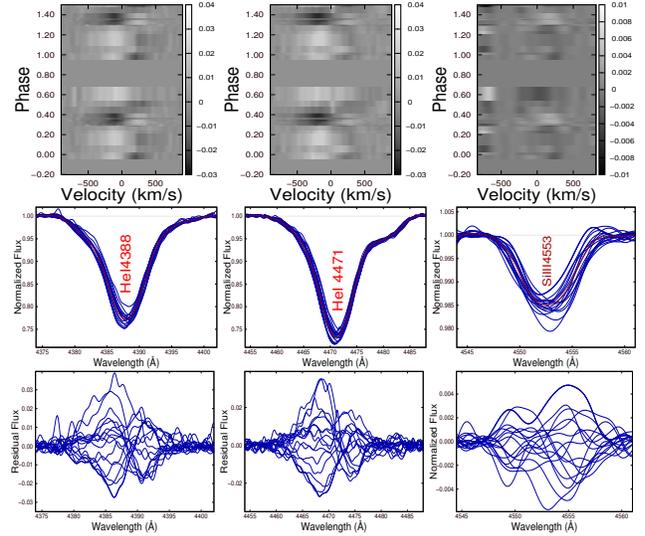}
\caption{
Same as in Fig.~\ref{fig:hydr}, but for the helium lines \ion{He}{i}~4388, 
\ion{He}{i}~4471, and the silicon line \ion{Si}{iii}~4553.
}
\label{fig:he}
\end{figure}

\begin{figure}
\centering
\includegraphics[width=0.47\textwidth] {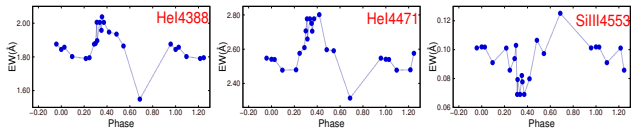}
\caption{
Variability of EWs of He and Si lines in FORS\,2 spectra obtained at eighteen different rotational phases.
}
\label{fig:ew_he}
\end{figure}

In Fig.~\ref{fig:he}, we present the overplotted profiles,
the differences between the individual and average profiles, and 
the differential dynamic plots
for the helium lines \ion{He}{i}~4388 and 
\ion{He}{i}~4471 and the only sufficiently strong silicon line
detected in the low-resolution FORS\,2 spectra, \ion{Si}{iii}~4553.
Distinct differences are detected in the behaviour between the two elements: He absorption lines 
are red-shifted in the phase ranges from about 0.55 to 0.70, around the phase 0, and from 0.1 to 0.2.
In the  phase 0.3-0.4, He lines and the silicon absorption line \ion{Si}{iii}~4553 are blue-shifted.
The offsets to the blue
and to the red are indicative of the presence of surface He and Si spots similar to the finding
of He and Si spots on the surface of $\sigma$\,Ori\,E \citep{Reiners2000}.
The results of the analysis of the variability of EWs of He and Si lines support the presumption of 
the presence of an inhomogeneous He and Si distribution.
As is shown in Fig.~\ref{fig:ew_he}, the Si line strength increases
in the phase range from 0.5 to 0.7, while the intensity of the He lines decreases in the same phase range. For both 
elements we do not detect any clear correlation with the location of the magnetic poles.
The error bars of all presented EW measurements are of the order of the symbol size and less.

\section{Discussion}
\label{sect:disc}
Our spectropolarimetric monitoring using FORS\,2 at the VLT of the rigidly rotating magnetosphere
star HD\,345439 revealed the presence of a strong magnetic field with a minimum polar
strength of 9\,kG reversing over the very short
rotation period of 0.77018\,d. Both the dipole strength and the very short 
rotation period of this star are similar to those
discovered in two other stars, HR\,5907 and HR\,7355 with half-day rotation periods
\citep{Grunhut2012,Rivinius2013}, known to belong to 
the group called the $\sigma$\,Ori\,E analogs (e.g.\ \citealt{Groote1997}). Apart from
HD\,345439, \citet{Eikenberry2014} identified another rigidly rotating magnetosphere
star, HD\,23478, rotating with a period of 1.04\,d \citep{Jerzykiewicz1993} and a strong
kG magnetic field \citep{Hubrig2015a,Sikora2015}. Among the four known
very fast rigidly rotating magnetosphere stars,
three of them, HD\,345439, HD\,23478, and HR\,5907, show   
low obliquities of their magnetic axes. For these stars it is expected that the plasma clouds are located 
close to the equatorial plane (e.g. \citealt{Townsend2005}).
Due to the presence of strong kG magnetic fields and fast rotation, such stars can serve as excellent 
laboratories to study the magnetic field and element abundance distributions using Zeeman Doppler Imaging, as well as
the effect of the magnetic field configuration on the angular momentum loss and the 
associated spin-down.

The study of the variability of the He and Si lines showed the presence of significant chemical
abundance variations across the stellar photosphere. However, no clear correlation with the position of the 
magnetic poles is indicated in our data.
Future high-resolution  high signal-to-noise spectropolarimetric observations will be worthwhile to determine
the locations of these abundance spots as well as the surface distribution of other elements.  

Variable emission wings, most clearly detected in the H$\beta$ line, become stronger at the rotational phase 
0, which corresponds to the best visibility of the positive magnetic pole.
The blue-shifted emission appears stronger and more extended than the redshifted emission. This behaviour,
which differs from the behaviour of the near-IR lines in HD\,345439, was already observed in a few other 
stars with magnetospheres (e.g., HR\,5907 -- \citealt{Grunhut2012}; HR\,7355 -- \citealt{Rivinius2013}; 
HD\,23478 -- \citealt{Sikora2015}; CPD\,$-$62$^{\circ}$\,2124 -- Hubrig et al., in preparation).
Due  to  the  shortness of the rotation periods and the presence  of very strong magnetic fields in the 
atmospheres of the  $\sigma$\,Ori\,E analogs,
these stars are the best candidates to carry out multiwavelength observations at different optical depths
to constrain their magnetospheres in more detail (e.g. \citealt{Carciofi2013}) and to study various atmospheric effects that
interact with a strong magnetic field.

\section*{Acknowledgments}
\label{sect:ackn}
We thank the anonymous referee for useful comments.
Based on observations obtained in the framework of the ESO Prg.\ 097.D-0428(A).
AK acknowledges financial support from RFBR grant 16-02-00604A.
We would like to thank J.~R.~Lomax for valuable comments on the manuscript.

\label{lastpage}

\end{document}